  \documentstyle[aps,prb,multicol,epsf]{revtex}
  
\begin{document}
\renewcommand{\topfraction}{0.95}
\renewcommand{\bottomfraction}{0.95}
\renewcommand{\textfraction}{0.05}
% \input epsf.tex
% \tighten
\draft
\title{``Vortex-melting front'' in thin superconductors with pinning} 
\author{M.V.~Indenbom}
\address{Institute for Solid State Physics, 142432 Chernogolovka, 
       Moscow district, Russia \\
       Laboratoire des Solides Irradi\'es, CNRS UMR 7642 and
       CEA/DSM/DRECAM, Ecole Polytechnique,\\
       91128 Palaiseau cedex, France\\
       Max-Planck-Institut f\"{u}r
       Metallforschung, Postfach 800665, D-70506 Stuttgart,
       Germany }
\author{E.H.~Brandt}
\address{Max-Planck-Institut f\"{u}r
       Metallforschung, Postfach 800665, D-70506 Stuttgart,
       Germany}
\author{C.J.~van der Beek and M.~Konczykowski}
\address{Laboratoire des Solides Irradi\'es, CNRS UMR 7642 and
       CEA/DSM/DRECAM, Ecole Polytechnique,\\
       91128 Palaiseau cedex, France}
\date{Submitted to Physical Review B: May 8, 2002, resubmitted: August 23, 2002}
%for cond-mat
\maketitle

\begin{abstract}

Magneto-optical observations of a second flux front,
which occurs at the second peak in the magnetization of 
Bi$_2$Sr$_2$CaCu$_2$O$_x$
single crystals related to the known first order ``vortex-lattice
melting'', are reconsidered. We show that, in thin samples,
electrodynamics necessarily leads to an extended region in
which the magnetic induction adopts a nearly constant value close to
that at which the phase transition occurs at thermal equilibrium.
In this region a dynamical phase mixture of vortex ``solid'' and
``liquid'' should exist.
Interestingly, the observed second flux front does not mark the
``melting'' front, as it was naively interpreted earlier, but
indicates the disappearance of the last ``solid droplets''.

\end{abstract}
\begin{multicols}{2}
\narrowtext

 \section{Introduction}

One of the most fascinating features of the vortex structure in
layered superconductors is the so-called ``melting'' of the vortex
ensemble. It is clearly established by now that this transition
is a first-order phase transition above which the long-range correlation
along the vortex lines is lost.\cite{NatureMelting} The most
thoroughly investigated material, in which the ``vortex melting''
occurs well below the second critical field, is the high-$T_c$
superconductor Bi$_2$Sr$_2$CaCu$_2$O$_x$ (BSCCO). There are
attempts to add some other phase transition lines to the
$B$-$T$ phase  diagram of this material ($B$ is the local
induction and $T$ the temperature). For example, the field of
abrupt increase of the screening current with increasing $B$,
the so-called
second peak of the magnetization, was ascribed to a second order
phase transition connected with the ``melting'' line at a
tricritical point. However inspection of the
available data suggests
that the line of the second peak in the phase diagram is
the continuation of the ``melting'' line into the low-temperature
area where bulk currents do not completely relax during
the experiment.\cite{OxygenMelting}

It is natural to assume that, in spite of the
macroscopic irreversibility, the vortex phases are in equilibrium
locally. Then the unified ``melting line'' including
the second peak-line,
which shifts as one curve under the introduction of
additional disorder and/or change of anisotropy by oxygen doping
or electron irradiation, \cite{OxygenMelting,IrradiationMelting}
is the only first-order phase transition line.
Recently the absence of the tricritical point at the border of
observable pinning
was elegantly demonstrated by unmasking the equilibrium
magnetization step at the vortex melting point from the hysteretic
magnetization using ac demagnetization.
\cite{No3point}
This picture implies
different pinning properties of these two vortex phases with
correspondingly different current--voltage characteristics,
which were extracted from local Hall probe
measurements of the magnetic relaxation in the vicinity
of the second magnetization peak at different temperatures
and starting fields.\cite{Hall2Jc}
The resulting jump of the screening current at the first-order
phase transition is then observed as a second front of the
penetrating magnetic flux.

Magneto-optical observations of this second magnetization front
initially did not bring much new information and
did not attract enough attention. 
\cite{Indenbom2ndFront,Vlasko2ndFront} Such
experiments were initially interpreted straightforwardly as
the direct observation of the phase boundary
which was already well established by local Hall probe
measurements.\cite{Majer94,Hall2Jc} 
However, after more careful consideration it
became clear that the suggested ``naive'' interpretation is not
compatible with the (macroscopic) electrodynamics of thin samples.
Namely, in thin superconductors a sharp boundary between two
different screening currents should produce the same {\it
logarithmic peak} in the induction as at the sample edges. 
However, this moving peak {\it is not observed} in experiments. 
The correct theory indeed yields that a transition region with 
finite width and smooth current distribution should be expected, 
as it is known from the exact solutions for the penetration of
perpendicular flux into thin samples, see
{\em e.g.} \ Ref.~\onlinecite{StripEPL}.  
We propose here a model which predicts that this transition area
in thin samples should inevitably contain a
{\it mixture of both phases}.

A comprehensive description of the peculiarities of magnetic
flux dynamics became particularly important when new observations
of metastable vortex phases under fast remagnetization were
reported. \cite{MetaBarIlan,MetaPalaiseau}   In order to
separate new dynamic effects from the (already
complicated) geometry-dependent ``regular'' flux dynamics
occurring when the vortex phases are in local equilibrium, we
make a step backward in our investigations and
describe the equilibrium case more carefully. The nature of
flux flow at the phase transition is simulated qualitatively using an
algorithm developed for the flux dynamics in thin conductors
with nonlinear current-voltage law. \cite{EHBdynamics}

\section{Magneto-optical observations}

The penetration of a magnetic field into BSCCO single crystals
was visualized by means of the magneto-optical imaging technique
using a ferrimagnetic garnet indicator film with in-plane
anisotropy. \cite{MOpt1,MOptReview,ReviewJooss}
In general, higher light
intensity in the image means higher local magnetic induction normal
to the sample surface. The images were recorded by a CCD camera
during the continuous ramp of the perpendicular magnetic field
with $dB_a/dt = 200$ G\,s$^{-1}$ and digitalized every
250 ms by a video-frame grabber card.
The camera was working at the standard TV frame frequency of
25 Hz, so each image of the obtained movie is an average over
about 40 ms. The increase of $B_a$ during the acquisition of a
single image is comparable to the measurement noise.
The relatively fast ramp of the field was used in order to
observe the critical slope of the initial penetration,
which otherwise relaxes. The intensity profiles measured across
the digital images were calibrated in terms of field units using an
additional recording under equivalent conditions but above $T_c$.
A separate calibration curve was obtained from the additional
``empty'' images for each point on the flux profiles.
In this way we reduce the effect of inhomogeneities in the illumination 
and in the sensitivity of the indicator across the image.
 For demonstration, an underdoped BSCCO crystal grown at the
FOM-ALMOS center (the Netherlands)
by the traveling-solvent floating zone
technique under 130 mbar O$_2$ partial pressure\cite{MingLi2002} 
was specially selected
for the uniformity of flux penetration and for the low field
$B_{\rm sp} \approx 180$ G at which the second magnetization peak
occurs. The crystal was cut into a rectangle with dimensions
$1450 \times 700 \times 43 \, \mu$m$^3$.

The initial penetration into the zero-field-cooled sample
resembles the numerous magneto-optical observations of the
Bean-like critical states in type-II superconductors. 
The crystal edges are well observed as a sharp change from the very
bright field outside the sample and the grey interior.
The flux penetrating from the crystal edges is visible as a grey belt
(higher brightness represents higher % to continue

%%%%%% FIGURE 1 %%%%%%%%%
\begin{figure}
\centerline{\epsfxsize 7cm \epsfbox{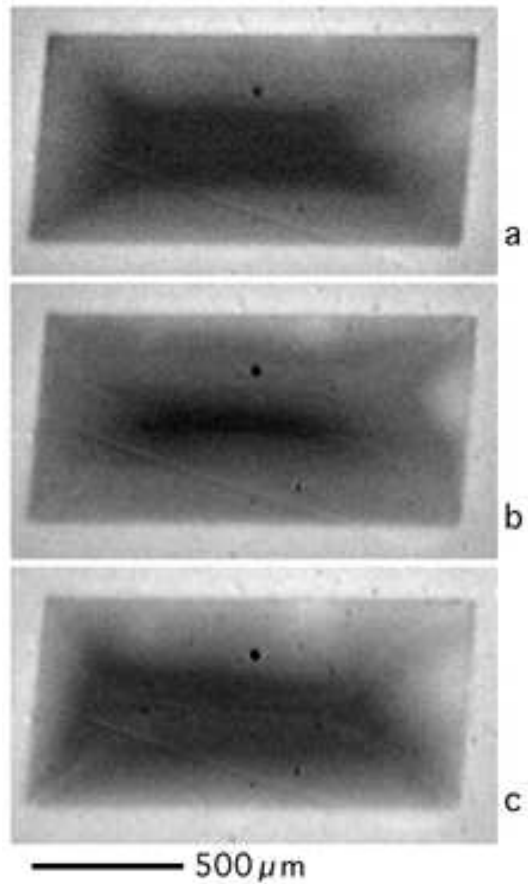}}
\caption{Magneto-optical imaging of the flux penetration into a
zero-field-cooled BSCCO single crystal at $T=28.8$~K in
continuously ramped $B_a$ with ramp-rate of 200~G\,s$^{-1}$.
In every image, light areas correspond to high induction,
while dark areas have low induction; however, the
intensity-to-induction mapping (look-up table)
in different images is selected separately
in order to reveal specific details. Selected snapshots from the movie:
(a)~$B_a = 200$~G, ordinary flux front below $B_{\rm sp}$. The dark
area in the center is the Meissner area with $B=0$, the bright area
around the crystal corresponds to the applied field $B_{a}$ and the grey belt
at the edges represents the classical Bean-like penetration with a
critical slope of $B$.
(b)~$B_a = 300$~G: the second flux front appears as a new, sharp increase
of the image brightness at the top and right sample edges,
while the initial field profile persists in the sample center.
(c)~$B_a = 450$~G: the second flux front above $B_{\rm sp}$.
Note that the dark central part now corresponds
to $B \sim B_{\rm sp}$, unlike the similar frame (a), where
this value is represented by a light grey tone.
The calibrated flux profiles for the entire image sequence,
shown in Fig.~2,  are taken along a vertical line crossing
the sample at its center downwards. }
\label{2ndFront}
\end{figure}
%%%%%% END FIGURE 1 %%%%%%%%

\noindent
local value of the perpendicular magnetic 
induction) which squeezes the central (dark
grey) Meissner area in which $B$ is zero (Fig.~\ref{2ndFront}~a).
The gradual decrease of the brightness of the penetrated perimeter
corresponds to the well-known Bean gradient of the penetrating flux.
When the field at the edges reaches $B_{\rm sp}$, a second sharp
flux front starts to penetrate from the crystal edges
(Fig.~\ref{2ndFront}~b).
\cite{Indenbom2ndFront,Vlasko2ndFront,MetaPalaiseau}
When the second flux front has penetrated deep into the sample,
the resulting image
is becoming nearly identical to the initial penetration 
(compare Fig.~\ref{2ndFront}~a and b).
However, one should remember that the induction in the
central dark area in Fig.~\ref{2ndFront}~c
approximately corresponds to that in the light grey field outside
the sample in Fig.~\ref{2ndFront}~a, where $B=0$ in the center.

%%%%%%%%%% FIGURE 2 %%%%%%%%%%
\epsfclipon
\begin{figure}
\epsfysize 16\baselineskip
\centerline{\epsfbox{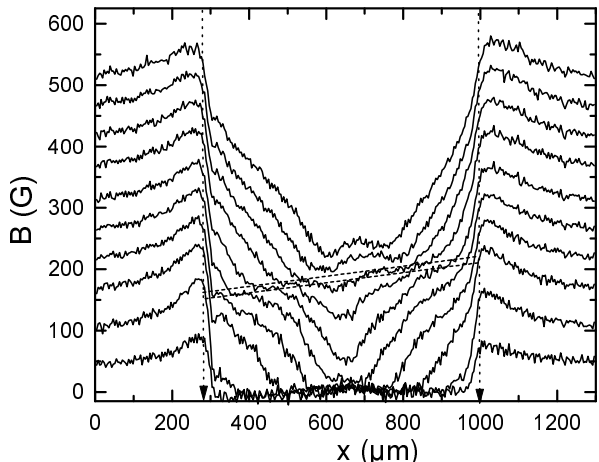}}
\caption{Magnetic flux profiles across the sample shown in Fig.1
at $T=28.8$ K. The field of the first order ``vortex melting'' is
indicated by two dashed lines. The position of the sample edges
is marked by vertical dotted lines with arrows at the abscissa.
The profiles represent a sequence of images taken every 0.25 s,
i.e., at intervals $\Delta B_a = 50$~G.  The images for the
profiles number 4, 6 and 9 (counting from the bottom) are shown
in Fig.1 a, b, and c, respectively.}
\label{Profiles}
\end{figure}
%%%%%%%%%% end FIGURE 2 %%%%%%%%%%

The whole course of flux penetration during field
ramping from zero to 500 G is visualized in Fig.~\ref{Profiles}
by $B(x)$ profiles measured across the sample width.
After reaching a well-determined
field level the increase of the induction $B$ pauses
and $B$ is stabilized until a new flux front approaches
which increases $B$ further.
The plateau of constant induction is associated with
the phase transition at $B_{\rm sp}$. The slight inclination of this
plateau, which is marked by two dashed lines, is due to inhomogeneity
of the crystal. To our knowledge, such inhomogeneity is found in
all BSCCO crystals and is inevitable (see \em e.g. \rm
Refs.~\onlinecite{MeltingVisual,MeltingTc}).
During the subsequent decrease of the field a similar feature
appears at the same line $B_{\rm sp}$ without
notable hysteresis in the transition field.
The second flux front looks
very similar to what the ordinary Bean-like flux
profile would have looked like if
some flux relaxation had occurred during a temporal halt of the field 
ramp.  Another close analogy exists with respect to flux penetration 
into a superconductor after field cooling.  That analogy, in 
particular, shows that the initial interpretation of the second front 
as a phase boundary separating regions with very different critical 
currents, is not compatible with the perpendicular geometry of the 
experiment.  Such a sharp border between two currents at a definite 
field is possible only in the longitudinal geometry, where it occurs 
at the junction of two different field gradients.  In perpendicular 
geometry, however, any flux front penetrating into a thin 
superconductor is accompanied by strong sheet currents {\it inside the 
screened area}.  \cite{StripEPL,TheussForkl}

\section{Model calculations and discussion}

\begin{figure}
    \epsfysize 20 \baselineskip
    \centerline{\epsfbox{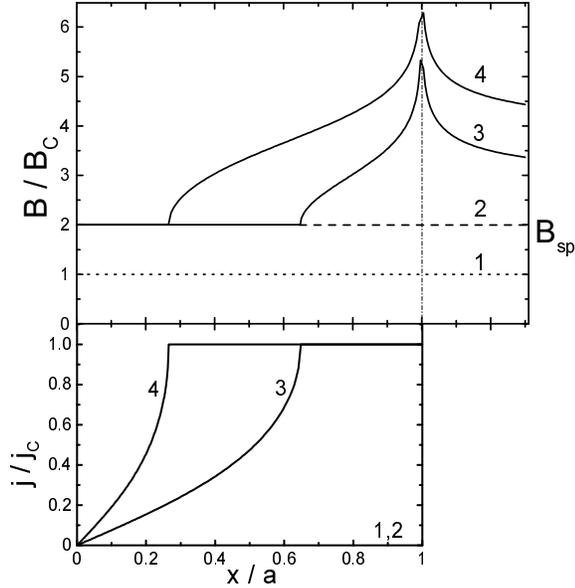}}
    \caption{A simplified model for the flux penetration in the
    presence of a first order vortex matter phase transition. the
    critical current is zero for $B < B_{\rm sp}$ and equal to 1 for $B >
    B_{\rm sp}$. Profiles 1 to 4 correspond to $B_{a} = B_{c}$, $2 B_{c}$,
    $3 B_{c}$, and $4 B_{c}$. The field $B_{\rm sp}$ is equal to $2 B_{c}$.}
    \label{fig:Bean}
    \end{figure}

In order to understand the experimental observations, let us first 
consider the simplest possible situation of a superconducting strip 
occupying  $|x| < a$, $|z| < d/2$ ($d \ll a$) in which 
the vortex matter undergoes a first order transition at 
$B_{z} = B_{\rm sp}$. 
In the vortex solid below $B_{\rm sp}$, the critical current is zero, 
and in the vortex liquid above
$B_{\rm sp}$, it has some non-zero value $J_{c}$.
In other words, the critical current discontinuously jumps from zero
to $J_{c}$ as the vortex matter melts. Furthermore, let us neglect 
surface and geometrical barriers. The flux penetration into this strip 
is illustrated in Fig.~\ref{fig:Bean}. 
As long as the applied field $B_{a} \parallel z$ does not exceed
$B_{\rm sp}$, the strip is perfectly transparent to the field 
(flat profiles 1 and 2 in Fig.~\ref{fig:Bean}). 
Once the applied field exceeds $B_{\rm sp}$, the internal induction is 
stabilized at the value $B_{\rm sp}$, and further flux penetration 
is screened by a non-zero $J(x)$. The distribution of this $J(x)$ 
and of the internal induction are described by the well--known 
solution of Ref.~\onlinecite{StripEPL}:

\[
\begin{array}{ccc}
    J(x) & = & \!\left\{
                  \begin{array}{ll}
                     \left( 2 J_{c}/\pi \right)
    \arctan \frac{cx}{ \sqrt{ b^{2} - x^{2} } },
                               &  ~~~~|x| < b \\
             J_{c} x /  |x|,   &  ~~~~b < |x|  < a
          \end{array}

          \right. \\

         &   & \\
    B(x) & = & \left\{

               \begin{array}{ll}
                     B_{\rm sp},  & ~|x| < b \\
             B_{\rm sp} + B_{c} {\rm arctanh} \frac{\sqrt{ x^{2} - b^{2}
             }}{c|x|},     & ~b < |x|  < a \\
             B_{\rm sp} + B_{c} {\rm arctanh} \frac{c|x|}{\sqrt{ x^{2}
             - b^{2} }},   & ~|x| > a
            \end{array}

        \right.

\end{array}
\]

\noindent 
where  $B_{c} = \pi \mu_{0} J_{c}/d$ is the characteristic field, 
$\pm b = a/\cosh [(B_{a}-B_{\rm sp})/B_{c}]$ 
is the position of the flux front,
and $c = \sqrt{a^{2}-b^{2}}$. For Fig.~\ref{fig:Bean}, we chose
$B_{c} = \frac{1}{2} B_{\rm sp}$. The situation is illustrated by the
profiles 3 and 4. In the region in which $B > B_{\rm sp}$, the current
$J = J_{c}$, while in the region with $B = B_{\rm sp}$, the strip geometry
imposes the smooth decrease of $J$ to zero in the interval between
the ``second'' flux front and the sample center. While it is evident
that the former region is occupied by the high field vortex liquid,
the latter region must also consist of vortex liquid, since the vortex
solid (with zero critical current) cannot sustain the non-zero
$J(x)$. Therefore, the observed flux front does not correspond to the
front of phase transformation.

The above consideration, based on the simple Bean model, does not
account for a non-zero current in the solid phase, nor for the
possibility of flux creep. We include these effects using the algorithm
previously developed for studying the flux penetration
into thin superconductors with a nonlinear current--voltage law
$E(J, J_{c}) = \rho_{0} |J / J_{c}|^{\sigma} J $ where $J$ is the
sheet current (the current density integrated over the
strip thickness). \cite{EHBdynamics}
We model the sharp jump of the current-voltage law from
%% %% between the critical currents
$E^{-}(J) = E(J,J^{-}_{c})$
in the low-field low-pinning phase ($J_{c} = J^-_{c}$)
to $E^{+}(J) = E(J,J^{+}_{c})$
in the high-field high-pinning phase ($J_{c} = J^+_{c}$)
by introducing a step in $J_{c}(B)$ at the transition field
$B_{\rm sp}$:

\begin{equation} % 1
  J_{c}(B) =J_{c0} +\frac{\Delta J_{c}}{2}
  \tanh\frac{B-B_{\rm sp}}{\Delta B} \,,
  \label{JcB}
\end{equation}

\noindent with jump width $\Delta B \ll B_{\rm sp}$ and jump height
$\Delta J_{c} =J^{+}_{c} -J^{-}_{c}$. Other models of steps in the
current--voltage characteristics, {\em e.g.} a jump in the
creep exponent $\sigma(B)$, or different values of the parameters,
give qualitatively similar results.

%%%%%%%%%%%% Additional figure 3 %%%%%%%%%%%%%%%%%%%%%%%%%
%%%%%%%%%%%%%% end Additional figure 3 %%%%%%%%%%%%%%%%%%%

\begin{figure}[b]
\epsfysize 39.4\baselineskip
\centerline{\epsfbox{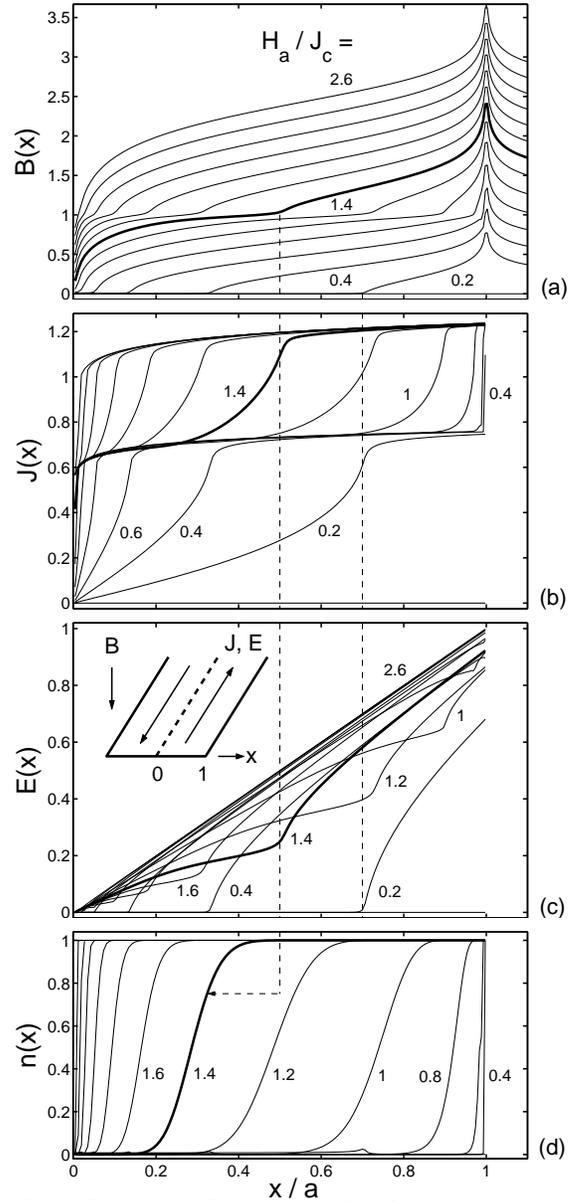}}
\caption{Computed flux penetration into a superconducting
strip with a sharp jump of the critical current from
$J^-_c = 0.75$ to $J^+_c = 1.25$
at the vortex-lattice phase-transition field $B_{\rm sp} = 1$:
(a) perpendicular induction $B(x)$,
(b) screening sheet current $J(x)$,
(c) electrical field $E(x)$, and
(d) relative concentration of the two phases $n(x)$
for $B_a = 0.2$, 0.4, $\dots$ 2.6.
The right dashed vertical line indicates the well-known initial
penetration front at $B_a = 0.2$ with a sharp decrease
of $J(x)$ from the critical value $J^-_{\rm sat} \approx 0.7$
and $E(x)$ dropping to zero.
To illustrate the correlation between $B(x)$,
$J(x)$, $E(x)$ and $n(x)$ during ``vortex melting'',
the profiles at $B_a=1.4$ are marked as bold lines. $B(x)$ clearly
exhibits a second flux front on top of $B_{\rm sp}=1$, starting
at the left vertical dashed line, where $J(x)$ sharply
decreases from $J^+_{\rm sat} \approx 1.2$
towards $J^-_{\rm sat} \approx 0.7$
and $E(x)$ has a sharp kink. But the corresponding
change of $n(x)$ extends much farther towards the center, as
indicated by an arrow pointing from the second front to the left.
The inset in (c) shows the geometry.
All values $B$, $J$ and $E$ are given in the dimensionless units
$J_c$ (see the text).
}
\label{Numerics}
\end{figure}

The following graphs (Fig.~\ref{Numerics}\,a--c) show profiles
(across the strip) of the magnetic field $B(x)$, screening current $J(x)$, 
and the electric field $E(x)$ under a constant ramp rate of the
external field $dB_{a}/dt = 1$ for $\sigma =20$, $B_{\rm sp}=1$,
$\Delta J_{c} = 0.5$ and
$\Delta B = 0.04$, in dimensionless units where $J^{+}_{c}=1$ (critical
sheet current), $a=1$ (half width of the strip), $B = H$,
$dE/dx = dB/dt$ (left-hand coordinate system), and $\rho_{0} =1$.
Our computed profiles of the induction $B(x)$
(Fig.~\ref{Numerics}a) well resemble the experimental profiles
(Fig.~\ref{Profiles}). The initial penetration at $B < B_{\rm sp}$
is the same as was calculated analytically \cite{StripEPL} and
numerically \cite{EHBdynamics}.
As usual, the corresponding $J(x)$ in the flux penetrated area saturates 
at the ``spatial projection'' $J_{sat}^{-}$ of the
low field current--voltage characteristic.
The reason for this is very simple: after the complete penetration
of the field, $B$ increases everywhere with the same constant rate.
From the Maxwell law $dE(x)/dx = dB(x)/dt$ it follows that
$E^-(J^-_{\rm sat}(x)) = x\,dB_a/dt \propto x$ everywhere.
Thus $J^-_{\rm sat}(x) =$ $J^-(x\,dB_a/dt)$, where $J^-(E)$ is the
inverse function of

%%%%%%%%%% FIGURE 3 %%%%%%%%%%
\epsfclipon
%%%%%%%%%% end FIGURE 3 %%%%%%%%%%

\noindent
 $E^-(J)$ defined 
below $B_{\rm sp}$. The constant critical current of the
classical Bean model corresponds to the nearly constant flat
part of $J^-_{\rm sat}(x) \approx J^-_{\rm sat}$. 
Unlike the Bean model,
in our more realistic model the abrupt change of direction 
of the critical current at $x=0$, the so-called discontinuity line 
(d-line),\cite{d_line} has a finite thickness.
The transition from $+0.8 J^-_{\rm sat}$ to $+0.8 J^-_{\rm sat}$
extends over $\Delta x_{d} \approx 2\,E^{-}(0.8\,J^-_c) / (dB_{a}/dt)$. 
For the particular $E(J)$ used in our simulations (Fig.~\ref{Numerics})
$\Delta x_{d} \approx 0.02 a$.

When $B$ increases further and approaches the
 value $B_{\rm sp}$, the field gradient decreases,
and further flux penetration occurs via the penetration of a new
flux front moving on top of the line $B \approx B_{\rm sp}$.
It is important to note that, as in the simplified case outlined above,
the screening current $J(x)$
does not jump sharply at the flux front but changes smoothly from
$J^-_{\rm sat}(x)$ to the new saturation
$J^+_{\rm sat}(x) \approx J^+_{\rm sat}$ {\it before} the
second front has arrived (Fig.~\ref{Numerics}b). Here again
$E^+(J^+_{\rm sat}(x)) = x\, dB/dt \propto x$ yields the
``spatial projection'' $J^+_{\rm sat}(x)$ of the high-field
current-voltage law. This behavior is valid for any pair of
functions $E^-(J)$ and $E^+(J)$, which play the role of
saturation profiles of the sheet current $J(x)$ after the
complete penetration of the field, below and above $B_{\rm sp}$,
correspondingly. Preceding the second front there is a region
where the flux density is nearly constant in space and time,
$B(x,t) \approx B_{\rm sp} $ (Fig.~\ref{Numerics}\,a);
in this region the slope of the electric field is reduced since
$ dE/dx = dB/dt $ is small. In order to reach its final saturation
shape $E(x) = x\, dB_a/dt$, $E(x)$ exhibits a sharp upward bend
at the second flux front, which is clearly  seen in
Fig.~\ref{Numerics}c. We should mention that a non-zero $\Delta B$ is
included here only for the stability of the numerical algorithm; its value
only affects the rounding of sharp features in the flux profile for
$B \approx B_{\rm sp}$, but not, as we checked, the wide smooth
transition of the current. The simplified solution outlined above
shows that the smooth transition in $J$ exists for $\Delta B = 0$ as
well.

Within the framework of our assumption that only two
phases with fixed current--voltage laws $E^-(J)$ and
$E^+(J)$ can exist in the superconductor, the above outlined
smooth transition of the resulting $J(x)$ between the
corresponding saturation profiles can be explained only by a
{\it phase mixture}. One can imagine various configurations of
the phase domains corresponding to different conditions,
varying from the equality of $E$ in the two phases to the homogeneity
of $J$. In general the dynamics of the corresponding phase mixture should
be solved, but this complication is outside the scope of our paper.
Here, we present the simplest and most natural model in which
the intermediate region consists of narrow
lamellar domains of alternating phases oriented parallel to the
flux flow, \em i.e.\rm perpendicular to the current, across the width
of the strip. This model corresponds to homogeneous $J$ across the phase
boundaries and a modulated $E$.
In the mixed region,  we then have $J_{\rm liquid} = $ $J_{\rm solid}
=$ $J(x)$ and $E(x)=n E^+(J(x)) + (1-n) E^-(J(x)) $. Here,
$n=n(x)$ is the relative concentration of one phase; $n=0$
corresponds to the ``vortex solid'' and $n=1$ to the ``vortex
liquid''. From the calculated $J(x)$ and $E(x)$ we obtain
$n(x)$ in the form

  \begin{equation}  % 2
  n(x) = \frac{E(x)-E^-(J(x))}{E^+(J(x))-E^-(J(x))}
  \label{Nx_ff}
  \end{equation}

\begin{figure}
\epsfysize 16\baselineskip
\centerline{\epsfbox{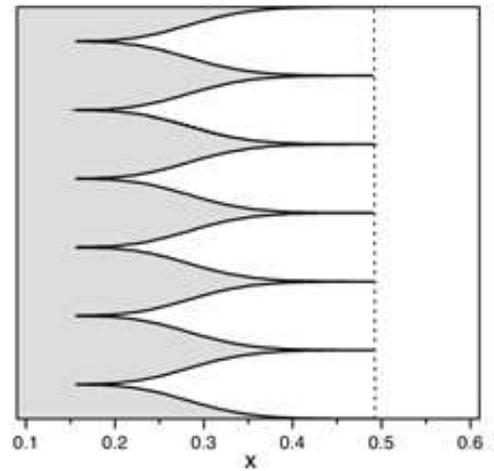}}
%% the final high resolution figure exist as Fig4.tif
\caption{Scheme of the phase lamella preceding the second flux
front marked here by a vertical dotted line that corresponds to
the left dashed line in Fig.~3.
The front moves from the right to the left.
 The ``liquid phase'' (right, white)
interpenetrates the ``vortex solid'' (left, light grey).
The vertical scale is not given by our theory.}
\label{Lamellas}
\end{figure}

\noindent (see Fig.~\ref{Numerics}d). 
In case of the simplest domain structure of the phase mixture, 
equal parallel lamellae along the flux flow, the relative width of 
the ``liquid'' domains is equal to $n(x)$.  
An example of such domains at 
$B_a=1.4$ is shown in Fig.~\ref{Lamellas}, which is obtained by 
flipping the corresponding profile $n(x)$ (outlined by the bold line 
in Fig.~\ref{Numerics}c) several times and cutting it at the right at 
the position of the second flux front, and at the left, {\em e.g.} at 
a value $n(x) = 0.01 $ ($0.01$ is an arbitrary choice, the final 
result depends on it logarithmically).

%%%%%%%%%% FIGURE 4 %%%%%%%%%%
\epsfclipon
%%%%%%%%%% end FIGURE 4 %%%%%%%%%%

Interestingly, ``melting'' ({\em i.e.} $n(x) > 0$) starts well before the
second flux front has arrived, and the second flux front
corresponds not to the midpoint $n=\frac{1}{2}$, but to the point
where $n(x)$ has already well saturated to $n=1$.
At this point the ``melting'' transition is complete.
The smooth transition of the current is characterized by the fast flux flow
through the weaker ``solid'' lamella  (light grey),
the width of which plays the role of a ``flux valve'',
while the vortices in the ``liquid'' lamella (white) are nearly
immobile until the second flux front arrives.
At $x \lesssim x_{\rm sp}$, the {\em average} electric field equals

\begin{displaymath}
E(x) \approx E_{\rm sp} = \int_{0}^{x_{\rm sp}} { dB(x,t) \over dt}
                          dx \approx {\rm const}
\end{displaymath}

\noindent because $B(x,t)$ is fixed to a
value $\sim B_{\rm sp}$. The vortex flow through the region
of the phase mixture is concentrated in the ``solid'' phase. The local
electric field (vortex velocity) in the solid is very high,  $E_{\rm
solid}=E^{-}(J(x))=E_{\rm sp}$, because the electric field  in the
``liquid'' phase $E_{\rm liquid}=E^+(J(x))$ is nearly zero. Consequently,
the width of the ``solid'' valve adapts itself so that the necessary
average flux flow is conserved, i.e. $1-n_{\rm sp} \approx
E_{\rm sp}/E_{\rm solid}$.

Particularly impressive is the moment when the second front
approaches and
the high second critical current $J^+_{\rm sat}$ passes through the
``weak'' ``solid'' phase. Here all necessary flux flow
squeezes into the very narrow ``solid neck'' with a huge speed
corresponding to $E^-( J^+_{\rm sat})$.
The order of the introduced values $E_{\rm liquid}$ and
$E_{\rm solid}$ can be read from the ``spatial projections''
of $E^+(J)$ and $E^-(J)$ in Fig.~\ref{Numerics}b
(see the corresponding lower and upper asymptotics
$J^+_{\rm sat}(x)$ and $J^-_{\rm sat}(x)$). For this
one has to take a horizontal projection of, for example, the
cross-point of the bold line $J(x)$ (at $B_a = 1.4$) and the
vertical dashed
line, indicating the second flux front, onto
$J^+_{\rm sat}(x)$ and $J^-_{\rm sat}(x)$. Then the first one
will give $x^+ = E_{\rm liquid} \sim 0.05$ and the second one
will be well outside the graph at the right
$x^- = E_{\rm solid} \sim 10^3$. The relative width of the
``solid neck'' is here $1-n_{\rm sp} \sim 2 \times 10^{-4}$.
In  reality, even for such domains as thick as the sample 
the ``neck'' width given by our qualitative theory
is well below the vortex spacing of 0.3 $\mu$m at $B_{\rm sp}$.
This means that probably only a single line of ``solid''
vortices will move before the second front arrives.

The proposed dynamical intermediate structure of the coexisting
phases should not be confused with the ordinary intermediate phase
that exists at any first order magnetic phase transition in the
narrow field range $\Delta B = B^+ - B^-$ over which the jump
occurs. The relative size of the equilibrium domains $n_{\rm eq}$
in that well known {\it static} case is fixed by the condition
that the average local field equals the $B(x)$ which is
determined by the entire current distribution,
$n_{\rm eq}B^+ +(1-n_{\rm eq})B^- =B(x)$.

As opposed to this, in our case the relative phase ratio $n(x)$
is determined by the different {\it dynamics} of the flux flow
in the two different phases and by the global electrodynamics of
the system, see the profiles in Fig.~\ref{Numerics}d and Eq.~(2).
In the ordinary static case, the region of the coexisting phases
may be very narrow because  $\Delta B$ is much smaller than
the typical variation of $B(x)$ over the sample; its width
equals $ (dB/dx)^{-1} \Delta B$ at the phase transition.
In our case the width of the transition region containing the
phase mixture is a sizable fraction of the sample size,
while our dynamic $B(x)$  varies even more due to the
presence of high bulk currents.

The question arises whether the intermediate structure of dynamically
coexisting solid and liquid vortex phases also occurs in thick
samples (thicker than the penetration depth $\lambda$). In the case of
the simplified model with zero current for $B < B_{\rm sp}$ and a critical
current $J_{c}$ for $B > B_{\rm sp}$, we refer to
Ref.~\onlinecite{thickBrandt}, in which it is shown that the shape of
the flux front in the interior of a thick superconductor is very well described
by the shape of the current profile in an infinitely thin superconductor.
Using Fig.~\ref{fig:Bean}, we can thus immediately draw the \em current
\rm front inside a thick sample. The region between this front and
the sample surface carries a current $J_{c}$ and must therefore be in
the liquid state. From the Bean model only, there is no
objection to the inner core remaining in the solid state with
$J_{c} = 0$. In that
case, one would also have a coexistence of solid and liquid phases,
but now across the sample {\em thickness}. Such a scenario would
depend on the stability of the curved phase boundary --- 
any phase mixture in the inner core up to its complete transformation
to the solid phase also does not contradict to electrodynamics.

%%%%%%%%%%%%%%%%% FIGURE 5 %%%%%%%%%%%%%%%%%%%%%%%%%%
\begin{figure}
    \vspace{1mm}
    \epsfxsize 6cm \centerline{\epsfbox{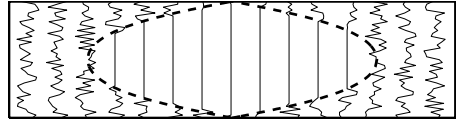}} \vspace{5mm} 
    \caption{Hypothetical phase front between the vortex solid and the 
    vortex liquid in a thick sample in the situation where the solid 
    cannot sustain any current flow, and the liquid has a non-zero 
    critical current.  The shape of the front just corresponds to that of 
    the current profile 3 in the lower panel of 
    Fig.~\protect\ref{fig:Bean}.  }
    \label{fig:thick}
\end{figure}
%%%%%%%%%%%%%%%%% END FIGURE 5 %%%%%%%%%%%%%%%%%%%%%%%%%%

Recently, a real-time differential treatment of
magneto-optical images allowed to visualize the propagating vortex
``melting'' front through a BSCCO crystal as a contour of the
related equilibrium magnetization jump.
\cite{MeltingVisual,MeltingTc}
The melting point appeared to be very sensitive to tiny
inhomogeneities of the
sample defect disorder and a very irregular ``vortex melting''
with no sign of
equilibrium domains is observed even in the best crystals.
Our lamella should be more robust because they are determined
by the critical currents, which are much higher than the tiny
equilibrium current drop at the
``vortex melting'' point.
But unfortunately, the elaborate procedure of such observations
cannot be applied to  our case of irreversible magnetization.
Very recently, after preparation of this paper, a
mixture of phases with different $E(J)$ has been observed in the peak
effect region in NbSe$_{2}$ using a scanning Hall probe magnetometer
and local ac excitation. \cite{domainsNbSe,VortexCapacitor} We have to
mention that the phenomenon observed there, is {\it static} and
the suggested model of the ``vortex capacitor'' differs from our
{\it dynamic} model. Furthermore, we deal with an {\it equilibrium
transition}, contrary to the {\it metastable} state considered in
Refs.~\onlinecite{domainsNbSe} and \onlinecite{VortexCapacitor}. 
The new observation
technique, nevertheless, may be applicable to our case; its magneto-optical
imaging counterpart can be realized as well. Note however that in our case
the picture is evolving rapidly (on the time scale of 10$^{-2}$ s) and the
irreversible currents which would
screen the small ac excitation are much higher.

\section{Conclusions}

We have obtained experimental and numerical flux profiles that render
the major features of flux flow at ``vortex-lattice melting'' in the
presence of pinning and magnetic hysteresis. The flux profiles show
a second front corresponding to the well-known second peak in the
magnetization curve. In thin superconducting plates, ``melting'' occurs
{\it well before} the arrival of the  second flux front.
At that moment, the phase transition is complete and the
low-field phase (``vortex solid'') with lower $J_c$ completely
disappears there. The appearance of the high field phase (``vortex
liquid'') does not produce any sharp feature on the observable
field profiles. Rather, there is a large region preceding the second
flux front in which domains of ``solid'' and ``liquid'' coexist
{\it dynamically\/}: the relative width of these domains is
controlled by the electrical field induced by the vortices
moving under the force exerted by the current, 
which flows continuously across the domains; 
the current in its turn is controlled by the overall electrodynamics 
of the system.  This intricate feature should be taken into account 
in a correct interpretation of experiments.

Evidently this effect occurs for any flux dynamics in the presence
of a first order phase transition, irrespective of whether it is
driven by ramping the applied field as in the presented experiment,
or it happens during magnetic relaxation
as in Ref.~\onlinecite{MetaBarIlan,MetaPalaiseau}.
Moreover, when the process is slow and the current in the ``weak''
``solid vortex-lattice'' phase is allowed to relax to zero, the
``melting'' proceeds nearly immediately everywhere because the
flux flow required for this is very low. In this case the remaining
``solid'' lamella should be very narrow and the second front
moves on top of the nearly homogeneous ``liquid''
without any relation to the phase transition itself.

The authors are grateful to Ming Li and P.H. Kes
of the FOM-ALMOS center (the Netherlands)
for the samples.
M.V.I.  \footnote{Contact author: \\ 
Mikhail Indenbom, e-mail: indenbom@issp.ac.ru} 
acknowledges financial support by the Max-Planck-Gesellschaft.
%\\

%
\end{multicols}
\end{document}